\documentclass[epj]{webofc}
\usepackage[varg]{txfonts}   
\usepackage{amsmath}
\usepackage{epsfig}
\usepackage{graphicx}
\usepackage{slashed}

\def\beq{\vspace{-0.3em}\color{magenta}\begin{eqnarray*}}
\def\eeq{\end{eqnarray*}\color{blue}\vspace{-0.3em}}
\def\bea{\begin{eqnarray}}
\def\eea{\end{eqnarray}}
\def\be{\begin{equation}}
\def\ee{\end{equation}}

\def \cf {C_F}
\def \nc {N_c}

\def\als{\alpha_{\rm s}}

\def\siml{{\ \lower-1.2pt\vbox{\hbox{\rlap{$<$}\lower6pt\vbox{\hbox{$\sim$}}}}\ }}     
\def\simg{{\
    \lower-1.2pt\vbox{\hbox{\rlap{$>$}\lower6pt\vbox{\hbox{$\sim$}}}}\ }}

\woctitle{International Conference on New Frontiers in Physics 2013}

\begin{document}

\title{Non-relativistic particles in a thermal bath}

\author{Antonio Vairo \thanks{\email{antonio.vairo@ph.tum.de}}}

\institute{Physik-Department, Technische Universit\"{a}t M\"{u}nchen, \\James-Franck-Str. 1, 85748 Garching, Germany}

\abstract{Heavy particles are a window to new physics and new phenomena. Since the late eighties they are treated
by means of effective field theories that fully exploit the symmetries and power counting typical of non-relativistic
systems. More recently these effective field theories have been extended to describe non-relativistic particles
propagating in a medium. After introducing some general features common to any 
non-relativistic effective field theory, we discuss two specific examples: 
heavy Majorana neutrinos colliding in a hot plasma of Standard Model particles in the early universe 
and quarkonia produced in heavy-ion collisions dissociating in a quark-gluon plasma.}

\maketitle

\section{Introduction}
\label{intro}
Heavy particles are a window to new physics for they may be sensitive 
to new fundamental degrees of freedom. Some of these new degrees of freedom may be 
themselves heavy particles (like, for instance, the heavy neutrinos that we will discuss in section~\ref{sec-Maj}).
Heavy particles can also be clean probes of new phenomena emerging 
in particularly complex environments. Examples are heavy quarks and quarkonia 
as probes of the state of matter formed in heavy-ion collisions.

We call a particle heavy if its mass $M$ is much larger than
any other scale $E$ characterizing the system. 
The scale $E$ may include the spatial momentum of the heavy particle, 
masses of other particles, $\Lambda_{\rm QCD}$, symmetry breaking scales, 
the temperature $T$ of the medium and any other energy or momentum scale 
that describes the heavy particle and its environment.
Under this condition the heavy particle is also non relativistic. 
The hierarchy $M \gg E$ calls for a low energy description 
of the system in terms of a suitable effective field theory (EFT) whose 
degrees of freedom are a field $H$ encoding the low-energy modes 
of the heavy particle, and all other low-energy fields of the system.
In a reference frame where the heavy particle is at rest up to fluctuations that are much smaller than $M$, 
the EFT Lagrangian has the general form 
\be
{\cal L} = H^\dagger i D_0 H + \hbox{higher-dimension operators} \, \times \hbox{powers of} \, \frac{1}{M}  + {\cal L}_{\rm light\; fields}\,.
\label{EFT1}
\ee
In the heavy-particle sector the Lagrangian is organized as an expansion in $1/M$.
Contributions of higher-dimension operators to physical observables are counted in powers of $E/M$.
It is crucial to note that the EFT Lagrangian can be computed setting $E=0$. 
Hence its expression is independent of the low-energy dynamics.
The prototype of the EFT \eqref{EFT1} is the heavy quark effective theory~\cite{Isgur:1989vq,Eichten:1989zv}.

In this contribution, we will concentrate on the case of a heavy particle of mass $M$ 
propagating in and interacting with a medium characterized by a temperature $T$ much smaller than $M$.
This is a special case of the previous one. The EFT Lagrangian that describes 
the system at an energy scale much lower than $M$ has again the structure \eqref{EFT1}.
As pointed out before, the temperature does not enter in the computation of the Lagrangian, which  
is fixed by matching at $T=0$. This means that the Wilson coefficients encoding the contributions 
of the high-energy modes can be computed in vacuum.

The temperature is introduced via the partition function of the EFT and affects 
the computation of the observables. Contributions of higher-dimension operators are counted in powers of $T/M$. 
In order to study the real-time evolution of physical observables and in particular 
decay widths, it is convenient to compute the partition function in the 
so-called \textit{real-time formalism}. This consists in modifying  
the contour of the partition function to allow for real time (see e.g.~\cite{LeBellac:1996}).
In real time, the degrees of freedom double. However, in the heavy-particle sector the second degrees of freedom 
decouple from the physical degrees of freedom so that, as long as loop corrections to light particles can be neglected, 
the only difference with $T=0$ EFTs consists in the use of thermal propagators ~\cite{Brambilla:2008cx}.

In the following, we will consider heavy particles interacting weakly with a weakly-coupled plasma.
We will compute for them the corrections to the width induced by the medium, 
which we call their \textit{thermal width}, $\Gamma$. 
In particular, in section \ref{sec-Maj} we will compute the thermal width of a heavy Majorana neutrino 
interacting weakly with a plasma of massless Standard Model (SM) particles in the primordial universe. 
Whereas in section \ref{sec-qua} we will compute the thermal width of a quarkonium, which, like the 
$\Upsilon(1S)$, is heavy enough to be considered a non-relativistic 
Coulombic bound state, and is produced in heavy-ion collisions 
of sufficient high energy that the formed medium is a weakly-coupled plasma of light quarks and gluons.

\section{Heavy Majorana neutrinos}
\label{sec-Maj}
We consider a heavy Majorana neutrino, described by a field $\psi$ of mass $M$ much larger 
than the electroweak scale, $M \gg M_W$, coupled to the SM only through Higgs-lepton vertices:
\be
\mathcal{L}=\mathcal{L}_{\hbox{\tiny SM}} 
+ \frac{1}{2} \,\bar{\psi} \,i \slashed{\partial}  \, \psi  - \frac{M}{2} \,\bar{\psi}\psi 
- F_{f}\,\bar{L}_{f} \tilde{\phi} P_{R}\psi  - F^{*}_{f}\,\bar{\psi}P_{L} \tilde{\phi}^{\dagger}  L_{f} \,,
\label{modelMaj}
\ee
where $\mathcal{L}_{\hbox{\tiny SM}}$ is the SM Lagrangian with unbroken SU(2)$_{\rm L}\times$U(1)$_{\rm Y}$ gauge symmetry, 
$\tilde{\phi}=i \sigma^{2} \, \phi^*$, with $\phi$ the Higgs doublet, $L_{f}$ are lepton doublets with flavor $f$, 
$F_f$ is a (complex) Yukawa coupling and $P_L = (1 - \gamma^5)/2$, $P_R = (1 + \gamma^5)/2$ 
are the left-handed and right-handed projectors respectively.
This extension of the SM provides a model for neutrino mass generation 
through the seesaw mechanism~\cite{Minkowski:1977sc,GellMann:1980vs}.
It also provides a model for baryogenesis through thermal leptogenesis~\cite{Fukugita:1986hr,Luty:1992un}.
For a recent review see~\cite{Drewes:2013gca}.

Let us consider baryogenesis. Differently from the SM, the model \eqref{modelMaj} has the potential 
to originate a large baryon asymmetry. The mechanism is the following. 
At a temperature $T \siml M$ the neutrino falls out of equilibrium.\footnote{
In \eqref{modelMaj} we have simplified the realistic case with more neutrino generations 
by considering only one heavy Majorana neutrino.
} 
This happens because, as the temperature decreases, recombination processes become less and less frequent 
while the neutrino decays in the plasma.
Since the neutrino is a Majorana particle, a net lepton asymmetry is generated. 
This is transferred to a baryon asymmetry through sphaleron transitions. 
The phases of the Yukawa couplings $F_{f}$ in the Lagrangian \eqref{modelMaj} provide 
extra sources of C and CP violations besides those in the SM.
Finally the generated baryon asymmetry is protected from washout after sphaleron freeze-out 
at a temperature $T\sim M_W$. Hence the model may fulfill 
the three necessary \textit{Sakharov conditions} for baryon asymmetry~\cite{Sakharov:1967} 
in a stronger way than the SM does and account for the observed baryon asymmetry in the universe.

\begin{figure}[ht]
\centering
\sidecaption
\includegraphics[width=4cm,clip]{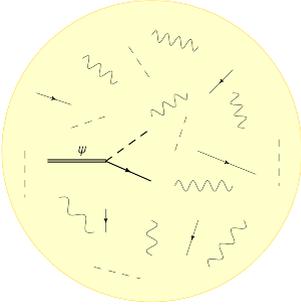}
\caption{Heavy Majorana neutrino decaying in the early universe plasma.}
\label{fig-early}   
\end{figure}

An important quantity for leptogenesis is the rate at which the plasma 
of the early universe creates Majorana neutrinos with mass $M$ at a temperature $T$.
This quantity is in turn related to the heavy Majorana neutrino thermal width in the plasma, 
see figure~\ref{fig-early}. We consider the temperature regime  $M \gg T \gg M_W$.
The heavy Majorana neutrinos are non-relativistic, with momentum 
\be
p^\mu = Mv^\mu + k^\mu, \qquad  k^\mu \ll M \,.
\ee
In kinetic equilibrium the residual momentum $k^\mu$ is of order $\sqrt{MT}$; far out of equilibrium it is of order $T$.
If the neutrino is at rest up to fluctuations that are much smaller than $M$, then $v^\mu = (1,{\bf 0})$.

At an energy scale much smaller than $M$ the low-energy modes of the 
Majorana neutrino are described by a field $N$ whose effective interactions with the SM 
particles are encoded in the EFT~\cite{Biondini:2013xua}:\footnote{
An EFT for heavy Majorana fermions has been considered also in~\cite{Kopp:2011gg}.}
\be
\mathcal{L} = \mathcal{L}_{\hbox{\tiny SM}}+\mathcal{L}_{\hbox{\tiny N}}\,,
\ee
where 
\be
\mathcal{L}_{\hbox{\tiny N}} = 
\bar{N} \left(i\partial_0 - \frac{i\Gamma_{T=0}}{2} \right)N 
+ \frac{\mathcal{L}^{(1)}}{M}
+\frac{\mathcal{L}^{(2)}}{M^2}
+\frac{\mathcal{L}^{(3)}}{M^3}
+\mathcal{O}\left(\frac{1}{M^4}\right)\,.
\label{EFT2}
\ee
The Lagrangian \eqref{EFT2} is of the type \eqref{EFT1}, the only difference being that 
the Majorana neutrino is a gauge singlet with a finite width at zero temperature, $\Gamma_{T=0}$, 
due to its decay into a Higgs and lepton. 

\begin{figure}[ht]
\centering
\sidecaption
\includegraphics[width=9.5cm,clip]{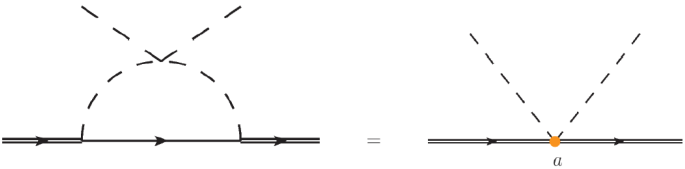}
\caption{One-loop matching condition for the neutrino-Higgs coupling.}
\label{fig-matching}   
\end{figure}

The power counting of the EFT indicates that the leading operators responsible 
for the neutrino thermal decay are dimension 5 operators contributing to $\mathcal{L}^{(1)}$.
The symmetries of the EFT allow only for one possible dimension 5 operator, which is 
\be
\mathcal{L}^{(1)}= a \; \bar{N} N \, \phi^{\dagger} \phi\,. 
\label{EFTL1}
\ee
This describes the scattering of Majorana neutrinos with Higgs particles.
Scatterings of Majorana neutrinos with gauge bosons, leptons or quarks are subleading.
The Wilson coefficient $a$ is fixed at one loop by the matching condition 
shown in figure \ref{fig-matching}. The left-hand side stands for an (in-vacuum) 
diagram in the fundamental theory \eqref{modelMaj}, whereas the right-hand side for 
an (in-vacuum) diagram in the EFT. Double lines are neutrino propagators, single lines 
lepton propagators and dashed lines Higgs propagators.
For the decay width only the imaginary part is relevant; it reads 
\be
{\rm{Im}}\,a  = -\frac{3}{8\pi}|F|^{2}\lambda\,,
\ee
where $\lambda$ is the four-Higgs coupling.

\begin{figure}[ht]
\centering
\sidecaption
\includegraphics[width=6cm,clip]{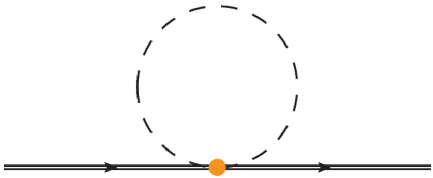}
\caption{Higgs tadpole contributing to the neutrino thermal width.}
\label{fig-loop}   
\end{figure}

The thermal width induced by \eqref{EFTL1} may be computed from the tadpole  diagram 
shown in figure~\ref{fig-loop}, where the dashed line 
has to be understood now as a Higgs thermal propagator.
The leading thermal width reads~\cite{Salvio:2011sf,Laine:2011pq},~\cite{Biondini:2013xua},~\cite{Bodeker:2013qaa}
\be
\Gamma = 2 \frac{{{\rm Im} \, a}}{M} \langle \phi^{\dagger}(0) \phi(0) \rangle_{T}  
=  -\frac{|F|^{2}M}{8\pi} \lambda \left( \frac{T}{M} \right)^{2}\,,
\label{leadingwidth}
\ee
where $ \langle \phi^{\dagger}(0) \phi(0) \rangle_{T}$ stands for the 
thermal condensate of the field $\phi$.
Note that in the EFT the calculation has split into a one-loop matching, shown in 
figure \ref{fig-matching}, which can be done in vacuum, and the calculation 
of a one-loop tadpole diagram, shown in figure \ref{fig-loop}, which is done in 
thermal field theory. The resulting simplification of the calculation with 
respect to a fully relativistic treatment in thermal field theory is typical 
of the EFT approach.

In a similar fashion one can calculate $T/M$ suppressed corrections to the thermal decay width.
Also for them the calculation splits into an in-vacuum matching of higher-dimension operators in the 
expansion \eqref{EFT2} and in the calculation of one-loop tadpole diagrams in thermal 
field theory. Only dimension 7 operators contribute to the width at next order in  $T/M$.
These are eight operators belonging to $\mathcal{L}^{(3)}$. Two each describe couplings of the 
Majorana neutrino to Higgs particles, leptons, quarks and gauge bosons respectively.
Finally, the thermal width at first order in the SM couplings and at order $T^4/M^3$ 
reads~\cite{Laine:2011pq},~\cite{Biondini:2013xua}
\bea
&& \Gamma 
=  \frac{|F|^{2}M}{8\pi}\left[-\lambda \, \left( \frac{T}{M} \right)^{2} 
+ \frac{\lambda}{2} \frac{{\bf k}^{\,2}\,T^2}{M^4}
- \frac{\pi^{2}}{80}\,(3g^{2}+g'^{\,2}) \,\left( \frac{T}{M} \right)^{4}
- \frac{7\pi^{2}}{60}\,|\lambda_{t} |^{2}\,\left( \frac{T}{M} \right)^{4} \right]\,,
\eea
where $g$ is the SU(2) coupling, $g'$ the U(1) coupling, and $\lambda_t$ the top Yukawa coupling.

\section{Heavy quarkonia}
\label{sec-qua}
\begin{figure}[ht]
\centering
\sidecaption
\includegraphics[width=4cm,clip]{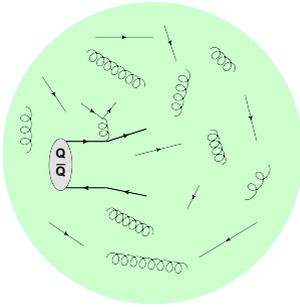}
\caption{Heavy quarkonium dissociating in a plasma of light quarks and gluons.}
\label{fig-hi} 
\end{figure}

Heavy quarkonia are bound states of heavy quarks. A quark is considered heavy if 
its mass $M$ is much larger than the typical hadronic scale $\Lambda_{\rm QCD}$. 
Quarkonia include bound states of charm and bottom quarks.
They are a probe of the state of matter made of gluons and light quarks 
formed in high-energy heavy-ion collisions~\cite{Matsui:1986dk}.
The reasons are that heavy quarks are formed early in heavy-ion collisions, 
$1/M \sim 0.1 ~\hbox{fm} \ll 1~\hbox{fm}$, hence heavy quarkonium formation is sensitive to the medium,  
and that the quarkonium dilepton decay provides a clean experimental signal. The dissociation of a heavy quarkonium in 
a plasma of quarks and gluons is sketched in figure~\ref{fig-hi}.

Under the condition $M\gg T$ and for quarkonia formed almost at rest in the laboratory 
rest frame, the mass $M$ of the heavy quark is the largest scale in the system;  
as in the general framework discussed in the introduction, we may consider 
quarkonia non-relativistic particles suitable to be described by non-relativistic 
EFTs of the type \eqref{EFT1}. Differently from the Majorana neutrino case,  
quarkonia are, however, composite systems  characterized by several internal energy scales, 
which in turn may probe thermodynamical scales smaller than the temperature.
Hence the situation is more complex than the one discussed in section~\ref{sec-Maj}.
The energy scales characterizing a non-relativistic bound state are  
the typical momentum transfer in the bound state, which is of order $Mv$, 
and the typical binding energy, which is of order $Mv^2$. 
The parameter $v \ll 1$ is the relative heavy-quark velocity. This is of order $\als$ for a Coulombic bound state.
We call these scales the \textit{non-relativistic scales}.
The non-relativistic scales are hierarchically ordered: $M \gg Mv \gg Mv^2$.
Effective field theories exploiting the non-relativistic hierarchy in vacuum have been reviewed in~\cite{Brambilla:2004jw}.
For a weakly-coupled plasma, a relevant thermodynamical scale, which is smaller than the temperature,  
is the Debye mass, $m_D$, i.e. the inverse of the screening length of the chromoelectric interactions. 
This is of order $gT$, hence we have that parametrically $T \gg m_D$.
We call the scales $T$, $m_D$ and possibly smaller scales the \textit{thermodynamical scales}.
For a discussion on the energy scale hierarchy in the case of 
$\Upsilon(1S)$ produced in heavy-ion collisions at the LHC we refer to~\cite{Brambilla:2010vq,Vairo:2010bm}.
For first experimental evidence of suppression of excited bottomonium states at the LHC 
we refer to~\cite{Chatrchyan:2011pe}.

\begin{figure}[ht]
\centering
\sidecaption
\includegraphics[width=7cm,clip]{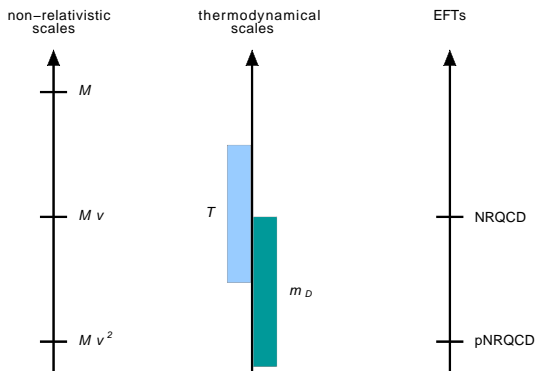}
\caption{Scales and EFTs for quarkonium in a thermal bath.}
\label{fig-EFTs}
\end{figure}

The existence of a hierarchy of energy scales calls for a description 
of the system in terms of a hierarchy of EFTs of QCD, which is the 
fundamental theory in this case. Many EFTs are possible in 
dependence of the specific ordering of the thermodynamical 
scales with respect to the non-relativistic ones. These are schematically shown in figure~\ref{fig-EFTs}.
For temperatures larger than those considered in figure~\ref{fig-EFTs} quarkonium does not form.

We call generically \textit{non-relativistic QCD} (NRQCD)~\cite{Caswell:1985ui,Bodwin:1994jh} 
the EFT obtained from QCD by integrating out modes associated with the scale $M$ 
and possibly with thermodynamical scales larger than $Mv$. 
The Lagrangian reads
\be
{\cal L} = \psi^\dagger \left( i D_0 + \frac{{\bf D}^2}{2 M} + \dots  \right) \psi   
+ \chi^\dagger \left( i D_0 - \frac{{\bf D}^2}{2 M} + \dots  \right) \chi  + \dots 
+ {\cal L}_{\rm light}\,,
\label{LNRQCD}
\ee
where $\psi$ ($\chi$) is here the field that annihilates (creates) the heavy (anti)quark.

We call generically \textit{potential NRQCD} (pNRQCD)~\cite{Pineda:1997bj,Brambilla:1999xf} 
the EFT obtained from NRQCD by integrating out modes associated with the scale $Mv$ 
and possibly with thermodynamical scales larger than $Mv^2$.  
The degrees of freedom of pNRQCD are  quark-antiquark states 
(cast conveniently in a colour singlet field S and a colour octet field O), 
low energy gluons and light quarks propagating in the medium.
The Lagrangian reads
\bea
{\cal L} &=& \int d^3r \; {\rm Tr}\, \left\{ 
   {\rm S}^\dagger \left( i\partial_0 -  \frac{{\bf p}^2}{M} -  V_s + \dots \right){\rm S} 
+  {\rm O}^\dagger \left( i{D_0}      -  \frac{{\bf p}^2}{M} -  V_o + \dots \right){\rm O}\right\}
\nonumber\\
&&
\hspace{7.1mm}
+ {\rm Tr} \left\{  {\rm O^\dagger}{\bf r}\cdot g {\bf E}\,{\rm S} + {\rm H.c.} \right\} 
+ \frac{1}{2} {\rm Tr} \left\{  {\rm O^\dagger}{\bf r} \cdot g {\bf E}\,{\rm O}+ {\rm c.c.} \right\} + \cdots
+ {\cal L}_{\rm light} \,,
\label{LpNRQCD}
\eea
where ${\bf E}$ is the chromoelectric field and $g$ is now the SU(3)$_c$ coupling.
Both EFT Lagrangians \eqref{LNRQCD} and \eqref{LpNRQCD} are of the form \eqref{EFT1}; 
only the field S is a gauge singlet. The pNRQCD Lagrangian is also organized as an 
expansion in $r$: $r$ is the distance between the heavy quark and antiquark, which is of 
order $1/(Mv)$. At leading order in $r$, the singlet field S satisfies a Schr\"odinger equation.
Hence the Wilson coefficient $V_s$ may be interpreted as the singlet quarkonium potential.
Similarly the Wilson coefficient $V_o$ may be interpreted as the octet quarkonium potential.
The explicit expressions of the potentials depend on the version of pNRQCD that is considered; 
in particular, if thermodynamical scales have been integrated out, $V_s$ and  $V_o$ 
may depend on the temperature. We have set equal to one possible Wilson coefficients appearing in the 
second line of \eqref{LpNRQCD}.

A key quantity for describing the expected quarkonium dilepton signal is the quarkonium dissociation width. 
At leading order it may be useful to distinguish between two dissociation mechanisms:
\textit{gluodissociation}, which is the dominant mechanism for $Mv^2 \gg m_D$, 
and \textit{dissociation by inelastic parton scattering}, which is the dominant mechanism for $Mv^2 \ll m_D$.
Beyond leading order the two mechanisms are intertwined and distinguishing 
between them would become arbitrary, whereas the physical quantity is the total decay width. 

\begin{figure}[ht]
\centering
\sidecaption
\includegraphics[width=6cm,clip]{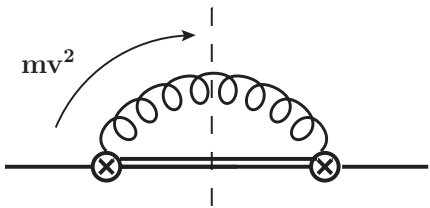}
\caption{Quarkonium gluodissociation in pNRQCD.}
\label{fig-gluodia}
\end{figure}

Gluodissociation is the dissociation of quarkonium by absorption of a gluon from the medium~\cite{Kharzeev:1994pz,Xu:1995eb}.
The gluon is lightlike or timelike (if it acquires an effective mass propagating through the medium). 
Gluodissociation is also known as \textit{singlet-to-octet break up}~\cite{Brambilla:2008cx,Brambilla:2011sg}.
The process happens when the gluon has an energy of order $Mv^2$.
Hence gluodissociation can be described in pNRQCD. In particular, it can be calculated by 
cutting the gluon propagator in the pNRQCD diagram shown in figure~\ref{fig-gluodia}, 
where the single line stands for a quark-antiquark colour singlet propagator, 
the double line for a quark-antiquark colour octet propagator and the  
circle with a cross for a chromoelectric dipole vertex.

\begin{figure}[ht]
\centering
\sidecaption
\includegraphics[width=6cm,clip]{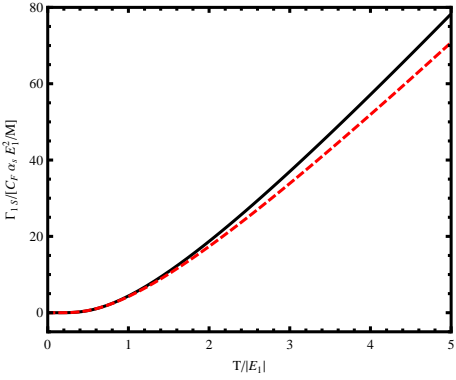}
\caption{$\Gamma_{1S}$ due to gluodissociation (continuous black line) vs the Bhanot--Peskin approximation (dashed red line).}
\label{fig-Gammagluo}
\end{figure}

Cutting rules at finite temperature~\cite{Kobes:1985kc,Kobes:1986za,Bedaque:1996af,Gelis:1997zv}  
constrain the gluodissociation width of a quarkonium with quantum numbers $n$ and $l$,  
$(Q\overline{Q})_{nl}$, which is at rest with respect to the medium, to have the form 
\be 
\Gamma_{nl}= \int_{q_\mathrm{min}}\frac{d^3q}{(2\pi)^3}\,n_{\rm B}(q)\,\sigma_{\rm gluo}^{nl}(q)\,,
\label{factgluo}
\ee
where $\sigma_{\rm gluo}^{nl}$ is the in-vacuum cross section $(Q\overline{Q})_{nl} + \hbox{gluon} \to Q + \overline{Q}$,  
and $n_{\rm B}$ is the Bose--Einstein distribution. The explicit leading order (LO) expression of 
$\sigma_{\rm gluo}^{1S}$ for a Coulombic $1S$ state like the $\Upsilon(1S)$ is~\cite{Brambilla:2011sg,Brezinski:2011ju} 
\be
\sigma^{1S}_{\rm gluo\, LO}(q)=\frac{\als\cf}{3} 2^{10} \pi^2  \rho  (\rho +2)^2 \frac{E_1^{4}}{Mq^5}
\left(t(q)^2+\rho ^2	\right)\frac{\exp\left(\frac{4 \rho}{t(q)}  
\arctan \left(t(q)\right)\right)}{ e^{\frac{2 \pi  \rho}{t(q)} }-1}\,,
\label{1Scoul}
\ee
where $\rho = 1/(\nc^2-1)$, $t(q) = \sqrt{q/\vert E_1\vert-1}$, $E_1 = - M\cf^2\als^2/4$ and 
$\cf = (\nc^2-1)/(2\nc)$. The corresponding width is shown in figure~\ref{fig-Gammagluo}. 
Note that the gluodissociation formula holds for temperatures such that $T \ll Mv$ and $m_D \ll Mv^2 \sim |E_1|$.
In figure~\ref{fig-Gammagluo} we also compare with a popular approximation, the so-called 
\textit{Bhanot--Peskin approximation}~\cite{Peskin:1979va,Bhanot:1979vb}. 
This is the large $\nc$ limit of the full result \eqref{1Scoul} (but keeping $C_F=4/3$ in the 
overall normalization). Taking the large $\nc$ limit amounts at neglecting final state interactions, i.e.  
the rescattering of a $Q\overline{Q}$ pair in a colour octet configuration.

\begin{figure}[ht]
\centering
\sidecaption
\includegraphics[width=10cm,clip]{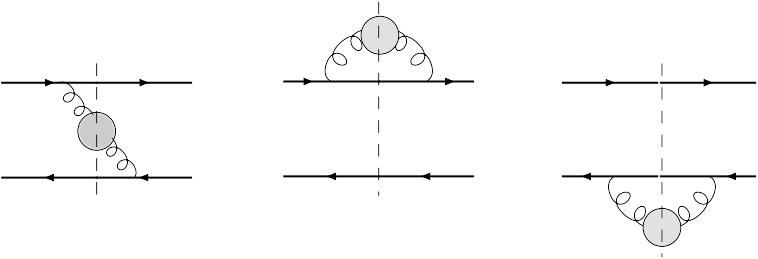}
\caption{Quarkonium dissociation by inelastic parton scattering in NRQCD.}
\label{fig-quasinrqcd}
\end{figure}

\begin{figure}[ht]
\centering
\sidecaption
\includegraphics[width=5.5cm,clip]{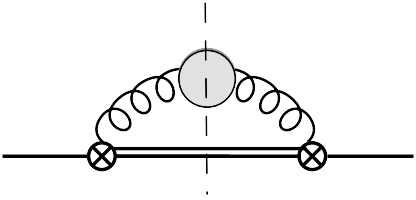}
\caption{Quarkonium dissociation by inelastic parton scattering in pNRQCD.}
\label{fig-quasipnrqcd}
\end{figure}

Dissociation by inelastic parton scattering is the dissociation of quarkonium by
scattering with gluons and light-quarks in the medium~\cite{Grandchamp:2001pf,Grandchamp:2002wp}.
The exchanged gluon is spacelike. Dissociation by inelastic parton scattering is also known 
as \textit{Landau damping}~\cite{Laine:2006ns}. Because external gluons are transverse,
according to NRQCD each external gluon is suppressed by $T/M$, 
see \eqref{LNRQCD}. At leading order, we may therefore just consider contributions 
to the width coming from cutting diagrams with a self-energy insertion
in one single gluon exchange~\cite{Brambilla:2013dpa}. If the exchanged gluon carries a momentum of 
order $Mv$, then the relevant diagrams may be computed in NRQCD, see figure~\ref{fig-quasinrqcd}.
If the exchanged gluon carries a momentum much smaller than $Mv$, 
then the relevant diagrams may be computed in pNRQCD, see figure~\ref{fig-quasipnrqcd}.
In both figures the dashed circle stands for a one-loop self-energy insertion.

\begin{figure}[ht]
\centering
\sidecaption
\includegraphics[width=6cm,clip]{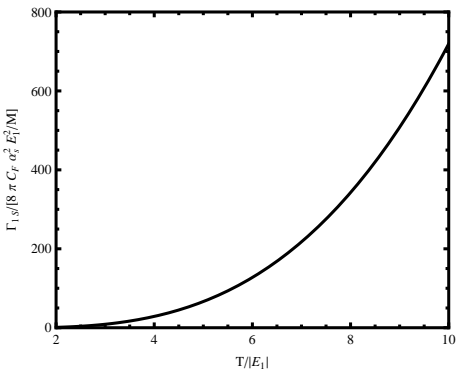}
\caption{$\Gamma_{1S}$ due to inelastic parton scattering. We have taken $m_Da_0 = 0.5$ and 
$|E_1|/m_D = 0.5$, where $a_0$ is the Bohr radius.}
\label{fig-Gammaparton}
\end{figure}

Cutting rules at finite temperature constrain the width by parton scattering 
of a quarkonium with quantum numbers $n$ and $l$, which is at rest with respect to the medium, 
to have the form 
\be
\Gamma_{nl}=\sum_p\int_{q_\mathrm{min}}\frac{d^3q}{(2\pi)^3}\,f_p(q)\,\left[1\pm f_p(q)\right]\,\sigma_p^{nl}(q)\,.
\label{factparton}
\ee
The sum runs over the different incoming and outgoing partons 
($p$ stands for parton, it may be either a light quark, $q$, for which the minus sign holds, 
or a gluon, $g$, for which the plus sign holds), 
with $f_g=n_{\rm B}$ and $f_q=n_{\rm F}$ ($n_F$ is the Fermi--Dirac distribution). 
The quantity $\sigma_p^{nl}$ is the in-medium cross section $(Q\overline{Q})_{nl} + p \to  Q + \overline{Q} + p$.
The convolution formula correctly accounts for Pauli blocking in the fermionic case (minus sign). 
Note that \eqref{factparton} differs from the corresponding 
gluodissociation formula~\eqref{factgluo} in the fact that it accounts for the thermal distributions 
of both the incoming and outgoing partons. Moreover, the cross section $\sigma_p^{nl}$ is not an 
in-vacuum cross section. Explicit expressions for the cross section in the case of 
a Coulombic $1S$ state like the $\Upsilon(1S)$ can be found in~\cite{Brambilla:2013dpa}.
These are valid for temperatures such that  $T \gg m_D \gg Mv^2 \sim |E_1|$.
The corresponding width is shown in figure~\ref{fig-Gammaparton}, where we have assumed 
three light quarks in the medium. Note the different normalization of the width with respect to
figure~\ref{fig-Gammagluo}.

\section{Conclusions}
\label{conclu}
In a framework that makes close contact with modern effective field theories 
for non-relativistic particles at zero temperature, one can compute 
the thermal width of non-relativistic particles in a thermal bath in a systematic way. 
In the situation $M \gg T$ one may organize the computation in two steps and compute 
the physics at the scale $M$ as in vacuum. If other scales are larger than $T$, then 
also the physics of those scales may be computed as in vacuum.
We have illustrated this on the examples of a heavy Majorana neutrino decaying in the early universe plasma 
and a heavy quarkonium dissociating in a weakly-coupled quark-gluon plasma.

\begin{acknowledgement}
I thank Simone Biondini, Nora Brambilla and Miguel Escobedo for collaboration on 
the work presented in section~\ref{sec-Maj} and  Nora Brambilla, Miguel Escobedo and 
Jacopo Ghiglieri for collaboration on the work presented in section~\ref{sec-qua}.
I acknowledge financial support from DFG and NSFC (CRC 110), and from the DFG cluster of
excellence “Origin and structure of the universe” (www.universe-cluster.de).
\end{acknowledgement}

\end{document}